\documentclass{ws-ijmpe}
\def\be{\begin{equation}}
\def\ee{\end{equation}}
\def\bea{\begin{eqnarray}}
\def\eea{\end{eqnarray}}

\def\eps{\varepsilon}
\def\eps0{\varepsilon_0}
\def\om{\omega}

\def\om{\omega}
\def\omd{\omega^{\dag}}
\def\OM{\Omega}

\def\eps{\varepsilon}

\def\<{\langle}
\def\>{\rangle}

\def\PLB{{\em Phys. Lett.}  B}

\def\PRC{{\em Phys. Rev.} {\bf C}}

\def\PTP{{\em Prog. Theor. Phys.}}

\def\sst{\scriptscriptstyle}

\def\be{\begin{equation}}
\def\ee{\end{equation}}
\def\bea{\begin{eqnarray}}
\def\eea{\end{eqnarray}}

\begin{document}

\markboth{R.Okamoto, S. Fujii and K. Suzuki}
{
Formal Relation among Various Hermitian and non-Hermitian
 Effective Interactions
}

%
\catchline{}{}{}{}{}
%

\title{
Formal Relation among Various Hermitian and non-Hermitian 
Effective Interactions
}

\author{\footnotesize Ryoji Okamoto}

\address{
Department of Physics, Kyushu Institute of Technology\\
Kitakyushu, 804-8550, Japan \\
okamoto@mns.kyutech.ac.jp}

\author{Shinichiro Fujii}

\address{
Department of Physics, University of Tokyo\\
Tokyo, 113-0033, Japan \\
sfujii@nt.phys.s.u-tokyo.ac.jp}

\author{Kenji Suzuki}

\address{
Senior Academy, Kyushu Institute of Technology\\
Kitakyushu, 804-8550, Japan \\
suzuki@mns.kyutech.ac.jp}

\maketitle
\begin{history}
\received{(received date)}
\revised{(revised date)}
\end{history}
\begin{abstract}
A general definition of the model-space effective interaction is given. 
The energy-independent
effective hamiltonians derived in a time-independent way
are classified systematically. 
\end{abstract}
\section{Introduction}

Nuclear many-body theory has been intensively studied for many years and 
reached a high level of understanding, but there still are some unresolved 
fundamental problems.

One of the leading approaches is to introduce an effective interaction and 
reduce the full many-body problem to a certain model-space problem.   
 There have been many review works concerning the effective interaction theory, for example, Ref. 1.

The principle of determining the effective interaction is that it should
have the property of decoupling between the model space and the excluded space
as discussed by Lee and one of the authors (K.S.).\cite{ref:SL80}   The property of decoupling is necessary for the effective interaction, 
but it does not give the condition of
determining uniquely the effective interaction.   Therefore many kinds of the 
effective interactions are possible.\cite{ref:SO83,ref:SO84a,ref:SO94}  Very recently Holt, Kuo and Brown\cite{ref:holt04} published a paper to study the 
various versions of  hermitian effective interaction.

The purpose
 of the present paper is to give a general definition of the effective interaction and show
that all the effective interactions in time-independent approach proposed so far can be classified 
systematically.
\section{Unified description of effective interaction}
We start with a general definition of an effective hamiltonian.   Let us write
the Schr\"odinger equation for a hermitian hamiltonian $ H $ defined in a full Hilbert space,
\be
\label{eq:scheq}
H |\Phi_k \> = E_k |\Phi_k\>,
\ee
where $|\Phi_{\sst k}\>$ is supposed to be orthonormal to each other,
\ $\langle \Phi_{k}|\Phi_{k'}\rangle =\delta_{kk'}.$

We introduce a model space that consists of some reference states.  We denote 
the projection operators onto the model space and its complement by $P$ and
$Q$, respectively.  
We consider an eigenvalue equation for a hamiltonian ${\cal H}$ and an overlap 
(or metric) $\chi$ defined in the $P$ space,
\be
{\cal H} | \xi_k \> = E_k \chi |\xi_k\>, \hspace{1cm} (k=1,2,\cdots,d),
\ee
where $d$ is the dimension of the $P$ space.   If $d$ eigenvalues 
$\{ E_k \}$ agree with the
eigenvalues of the original hamiltonian $H$, we call ${\cal H}$ the effective
 hamiltonian. If we separate the hamiltonian into the unperturbed hamiltonian
$H_{0}$, which has the property $H_{0}=PH_{0}P+QH_{0}Q$ and the perturbation (interaction), $V$, we call $V_{eff}={\cal H}-PH_{0}P$ the effective interaction.

We introduce an operator $\om$ that acts as
 a mapping between the $P$ and $Q$ spaces.\cite{ref:okubo} 
The operator $\om$ has the property, $\om = Q \om P$.
With the operator $\om$ we define an operator $X(n)$ given by
\footnote
{The definition of $X(n)$ is different from that given in the 
previous works.\cite{ref:SO94,ref:SO84a} The present definition is more
general than the previous one, but the effective interactions
to be derived are the same.
}
\be
\label{eq:xndef}
 X(n)=(1+\om-\om^{\dag})(1+\omd\om+\om\omd)^n,
\ee
where $n$ is an integer or a  half integer.
The inverse of $X(n)$ is given by
\be
\label{eq:xninv}
 X^{-1}(n)=(1+\omd\om+\om\omd)^{-n-1}(1+\omd-\om).
\ee
We note that the transformation $X(n)$ contains a special case
of $X(n=0)$ that is essentially equivalent to the transformation
 introduced by Navr\'{a}til, Geyer and Kuo.\cite{ref:NGK93}  
We refer to some important properties of $X(n)$ and $X^{-1}(n)$:
\begin{eqnarray}
\label{eq:xnp}
X(n)P=(P+\om)(P+\omd\om)^n,\\
QX^{-1}(n)=(Q+\om\omd)^{n-1}(Q-\om)
\end{eqnarray}
and
\be
\label{eq:qxxp}
QX^{-1}(m)X(n)P=0.
\ee

We consider a transformation of $H$ defined as
\be
\label{eq:hbarmn}
{\overline H}(m,n)= X^{-1}(m)H X(n),
\ee
and we also define an overlap operator
\begin{eqnarray}
\label{eq:kaimn}
\chi (m,n) &=& P X^{-1}(m) X(n) P \nonumber \\
           &=& (P + \omd\om)^{n-m}.
\end{eqnarray}
We now can prove the fact that if ${\overline H}(m,n)$ satisfies 
the following equation of decoupling
\be
\label{eq:qhbarp}
 Q{\overline H}(m,n)P=0,
\ee
which is explicitly written as\cite{ref:okubo}
\begin{equation}
\label{eq:qhp2}
 QHP+QHQ\om-\om PHP -\om PHQ \om = 0,
\end{equation}
the operator $P{\overline H}(m,n)P$ and $\chi (m,n)$
can be an effective hamiltonian and a 
corresponding overlap, respectively. The Eq.(\ref{eq:qhp2}) 
was also derived by 
 Poves and Zuker.\cite{ref:Poves-Zuker81}
%

We write the $P$-space effective hamiltonian for a set of $(m,n)$ as
\begin{eqnarray}
\label{eq:effhdef}
  {\cal H}(m,n) &=& P {\overline H}(m,n) P \nonumber\\
           &=& (P+\omd\om)^{-m-1}(P+\omd)H(P+\om)(P+\omd\om)^n.
\end{eqnarray}
The ${\cal H}(m,n)$ can also be expressed as
\be
\label{eq:calh}
  {\cal H}(m,n)=(P+\omd\om)^{-m}H(P+\om)(P+\omd\om)^n.
\ee
In the derivation of the above expression, we have used 
the relation written as
\be
  \omd H (P+\om)=\omd \om H (P+\om)
\ee
 or equivalently
 \be
 (P+\omd) H (P+\om)=(P+\omd \om )H (P+\om)
 \ee
which is obtained by multiplying the l.h.s. of Eq.~(\ref{eq:qhp2}) by
$\omd$.
\section{Classification of effective hamiltonians}
 
Among various effective hamiltonians we will discuss the detail of three 
kinds of the effective hamiltonians, namely,
${\cal H}(0,0),{\cal H}(0,-1)$ and ${\cal H}(-1/2,$\
$-1/2)$.
\begin{enumerate}
\item{$ m=n=0 $}

The effective hamiltonian ${\cal H}(0,0)$ is given by
\be
 \label{eq:h00}
 {\cal H}(0,0)=P H (P+\om),
\ee
and the corresponding overlap is $\chi (0,0)=P$.     The $ \om $ is related with the M${\o}$ller wave operator, $ \OM $, as
$ \Omega=P+\omega.$
The effective hamiltonian ${\cal H}(0,0)$ is written as
\begin{equation}
\label{eq:calh00}
 {\cal H}(0,0) = {\cal H}_{\sst {\rm B}} = P H \OM.
\end{equation}
This form has been used  as the standard effective hamiltonian
which was studied  by Bloch 
 and Horowitz.\cite{ref:BH58} 
   The structure
of ${\cal H}(0,0)$ would be the simplest, but it is non-hermitian.
  
\item{ $ m=0, n=-1 $}

The ${\cal H}(0,-1)$ is given explicitly by
\be
 \label{eq:h0-1}
 {\cal H}(0,-1)= P H (P+\om)(P+\omd\om)^{-1}.
\ee
The corresponding overlap is
\be
\label{eq:chi}
 \chi (0,-1)= (P+\omd\om)^{-1}.
\ee
The ${\cal H}(0,-1)$ given in Eq.~(\ref{eq:h0-1}) is apparently non-hermitian, 
but it is actually hermitian.   This is obvious from the expression of
 ${\cal H}(m,n)$ in Eq.~(\ref{eq:effhdef}).
The ${\cal H}(0,-1)$ is also written in a manifestly hermitian form as
\be
 {\cal H}(0,-1)
   =(P+\omd\om)^{-1}(P+\omd)H (P+\om)(P+\omd\om)^{-1}.
\ee
This effective hamiltonian agrees with 
Kato's effective hamiltonian\cite{ref:Kat49} defined originally as
\be
 \label{eq:kato}
 {\cal H}_{\sst K}= P {\overline P} H {\overline P}P
\ee
with the overlap 
$ \chi_{\sst K}= P {\overline P} P,$
where
$ {\overline P}
   =\sum_{\sst k=1}^{d}|\Phi_{\sst k}\>\< \Phi_{\sst k}|.$
  In terms of $\om$ we can prove\cite{ref:SO83}\hspace{1mm}that  
${\overline P}$ is written as 
\be
 \label{eq:pbar2}
 {\overline P}=(P+\om)(P+\omd\om)^{\sst -1}(P+\omd).
\ee
Substituting ${\overline P}$ in the above expression into Eq.~(\ref{eq:kato}), 
we readily see that ${\cal H}_{\sst K}$ is equivalent to
${\cal H}(0,-1)$.   Furthermore we also see, 
from Eq.~(\ref{eq:pbar2}), that $\chi_{\sst K}$ agrees 
with $\chi (0,-1)$ in Eq.~(\ref{eq:chi}).

\item{$ m=n=-1/2 $}

Using Eq.~(\ref{eq:effhdef}), ${\cal H}(-1/2,-1/2)$ is expressed as
\be
 \label{eq:h1212a}
 {\cal H}(-1/2,-1/2)
   =(P+\omd\om)^{\sst -1/2}(P+\omd)H(P+\om)(P+\omd\om)^{\ -1/2},
\ee
and from Eq.~(\ref{eq:calh}) we also have
\be
 \label{eq:h1212b}
 {\cal H}(-1/2,-1/2)
   =(P+\omd\om)^{\sst 1/2}H(P+\om)(P+\omd\om)^{\ -1/2}.
\ee
The above expression does not look manifestly hermitian, but 
 two expressions are
equivalent and ${\cal H}(-1/2,-1/2)$ is really hermitian.   
The corresponding overlap is
$\chi (-1/2,-1/2)=P$.

Historically several definitions of the hermitian effective 
hamiltonians have been proposed.   Among them we refer to the
following three hermitian effective hamiltonians, i.e.,
\begin{eqnarray}
\label{eq:calHp}
 {\cal H}_{\sst {\overline P}}
          &=& (P{\overline P}P)^{\sst -1/2}
                P{\overline P}H{\overline P}P
              (P{\overline P}P)^{\sst -1/2},\\
\label{eq:calHom}
 {\cal H}_{\sst \OM}
          &=& (\OM^{\dag}\OM)^{\sst 1/2}
                P H \OM (\OM^{\dag}\OM)^{\sst -1/2},\\
\label{eq:calHG}
 {\cal H}_G
          &=& P e^{-G} H e^{G} P.
\end{eqnarray}
  The exponent $G$ in Eq.~(\ref{eq:calHG}) is related with $\om$ 
as\cite{ref:Suz82a,ref:SR80,ref:Wes81}
\begin{eqnarray}
 \label{eq:zarc}
 G &=& {\rm arctanh} (\om-\omd), \  ( G^{\dag}=-G ) \nonumber\\
   &=& \sum_{\sst n=0}^{\sst \infty}
        \frac{(-1)^{\sst n}}{2n+1}\{ \om (\omd\om)^{\sst n}-{\rm h.c.}\}.
\end{eqnarray}
We can prove that all of ${\cal H}_{{\overline P}}$ ,
${\cal H}_{\OM}$ and ${\cal H}_G$ are equivalent to
${\cal H}(-1/2,-1/2)$ as follows:   Using Eq.~(\ref{eq:pbar2}) for
 ${\overline P}$,
we can easily show that ${\cal H}_{{\overline P}}$  and 
${\cal H}_{\OM}$ are equivalent to the expressions 
of ${\cal H}(-1/2,-1/2)$ in Eq.~(\ref{eq:h1212a}) and Eq.~(\ref{eq:h1212b}), respectively.
The ${\cal H}_G$ is of a 
canonical form, that is, ${\cal H}_G$ is derived by means of a unitary
transformation of the original hamiltonian $ H $.   We note here that
 $e^G $ with $ G $ given in Eq.~(\ref{eq:zarc}) 
becomes\hspace{1mm}\cite{ref:Suz82a} 
\be
 \label{eq:etoz}
 e^G = (1+\om-\omd)(1+\omd\om+\om\omd)^{\sst -1/2}
\ee
which is just equivalent to the transformation $X(-1/2)$, i.e.,
\be
\label{eq:xez}
X(-1/2)= e^G.
\ee
The ${\cal H}_{\sst G}$ is then written as
\be
 \label{eq:hzpxhxp}
 {\cal H}_{\sst G} =  P X^{\sst -1}(-1/2) H X(-1/2) P.
\ee
From the above expression the equivalence of ${\cal H}_{\sst G}$ to
${\cal H}(-1/2,-1/2)$ is obvious.
     The definition of  ${\cal H}_{{\overline P}}$
 is of des Cloizeaux\cite{ref:Clo60}\hspace{1mm}
 and ${\cal H}_{\OM}$ was
originally given by Okubo.\cite{ref:okubo}
 The structure of ${\cal H}_{\OM}$ was extensively studied 
by Brandow.\cite{ref:Bra67,ref:Bra75}    The 
canonical form ${\cal H}_G$ is often referred to 
as the Van Vleck form.\cite{ref:Vle29}
\end{enumerate}
\section{Family of hermitian effective interactions}
We re-start with the Schr\"odinger equation in a more concrete expression
\begin{equation}
\label{eq:s-eq-full-space}
(H_{0}+V)|\Phi_{k}\rangle = E_{k}|\Phi_{k}\rangle, \ (k=1,2,\cdots,d).
\end{equation}
The state vector $|\Phi_{k}\rangle$ is decomposed as
\begin{eqnarray}
|\Phi_{k}\rangle &=& P|\Phi_{k}\rangle +Q|\Phi_{k}\rangle 
   =|\phi_{k}\rangle +Q|\Phi_{k}\rangle
\nonumber\\
      &=& (P+\omega )|\phi_{k}\rangle, \ (|\phi_{k}\rangle =P|\Phi_{k}\rangle).
\end{eqnarray}
If we write the corresponding model-space Schr\"odinger equation as
\begin{equation}
\label{eq:model-space-eq}
 P(H_{0}+V_{LS})P|\phi_{k}\rangle =E_{k}|\phi_{k}\rangle,
\end{equation}
the non-hermitian effective interaction  $V_{LS}$ is expressed as 
\begin{eqnarray}
P V_{LS} P &=&P{\rm e}^{-\omega }(H_{0}+V){\rm e}^{\omega }P-PH_{0}P\nonumber\\
 &=& {\cal H}(0,0)-PH_{0}P.
\end{eqnarray}
We note that 
\be
P H \Omega P= P(H_{0}+ V_{LS})P.
\ee
The state vectors $|\phi_{k}\rangle$ are, in general, not mutually orthogonal,
since they are merely projections (onto $P$ space) of the orthogonal state vectors $|\Phi_{k}\rangle$. If we introduce the 
bi-orthogonal state, $|\tilde{\phi}_{k}\rangle$, corresponding to the model-space state $|\phi_{k}\rangle $ as 
$\langle \tilde{\phi}_{k}|\phi_{k'}\rangle =\delta_{kk'}$,
then the formal expression for  $\omega$ is given by
\begin{equation}
\label{eq:omega-formal}
\omega =\sum_{k=1}^{d}Q|\Phi_{k}\rangle \langle \tilde{\phi}_{k}|P.
\end{equation}
　Recently, Holt, Kuo and Brown\cite{ref:holt04} have used the $Z$ transformation method to study the various versions of hermitian effective interactions in order to 
reorient, or to suitably stretch,\cite{ref:brandow79}  
the vectors $|\phi_{k}\rangle$ such that they become orthonormal 
to each other.
\begin{eqnarray}
\label{eq:z-transformation}
Z|\phi_{k}\rangle &=& |v_{k}\rangle,\ (Z=PZP); \nonumber\\
 \langle v_{k} |v_{k'}\rangle &=& \delta_{kk'};(k,k'=1,2,\cdots,d).
\end{eqnarray}
Using the fact that the eigenvectors $|\Phi_{k}\rangle$ are orthonormal to each other we can derive 
\begin{eqnarray}
\delta_{kk'}&=&\langle \Phi_{k} |\Phi_{k'}\rangle 
            =\langle \phi_{k} |P+\omega^{\dag}\omega |\phi_{k'}\rangle
                   \nonumber\\
\label{eq:orthonormalization}
            &=&\langle v_{k} |(Z^{-1})^{\dag}(P+\omega^{\dag}\omega )
               Z^{-1}|v_{k'}\rangle.
\end{eqnarray}
Then we have 
\begin{equation}
\label{eq:p-omega-z}
P+\omega^{\dag}\omega =Z^{\dag}Z.
\end{equation}
The bi-orthogonal state $|\tilde{\phi}_{k}\rangle $ is related to 
$|v_{k}\rangle $ as
\begin{equation}
|\tilde{\phi}_{k}\rangle = Z^{\dag}Z | \phi_{k} \rangle
 = Z^{\dag}| v_{k}\rangle.
\end{equation}
One can easily check that 
$\langle \tilde{\phi}_{k} | {\phi}_{k'} \rangle
=  \langle {\phi}_{k} | Z^{\dag}Z |\phi_{k'}\rangle 
  =\langle v_{k} |v_{k'}\rangle =\delta_{kk'}.$
A formal expression for $\omega$ in Eq.~(\ref{eq:omega-formal}) 
is re-written as 
\begin{equation}
\omega = \sum_{k=1}^{d}Q|\Phi_{k}\rangle \langle v_{k}|Z.
\end{equation}
Using Eq.~(\ref{eq:z-transformation}) the model-space eigenvalue 
equation (\ref{eq:model-space-eq}) is transformed into
\begin{equation}
Z(H_{0}+V_{LS})Z^{-1}=\sum_{k=1}^{d}E_{k}|v_{k}\rangle \langle v_{k}|.
\end{equation}
Since $E_{k}$ is real and the vectors $|v_{k}\rangle $ are
 orthogonal to each 
other, $Z(H_{0}+V_{LS})Z^{-1}$ must be hermitian. Then the hermitian effective 
interaction, $V_{herm}$, is written as
\begin{eqnarray}
 V_{herm}&=& Z(H_{0}+V_{LS})Z^{-1}-PH_{0}P \nonumber\\
         &=& Z{\rm e}^{-\omega}(H_{0}+V){\rm e}^{\omega}Z^{-1}-PH_{0}P.
\end{eqnarray}
One must first obtain the $Z$ transformation in order to calculate 
$V_{herm}$.
There are certainly many ways to construct $Z$, and this fact generates a family of hermitian effective interactions, all originating 
from the non-hermitian
$V_{LS}$.\cite{ref:holt04}

One can construct $Z$ using the Schmidt orthogonalization procedure, 
and there are some ways in using the procedure.\cite{ref:holt04}
Holt, Kuo and Brown showed some well-known hermitization 
transformation.\cite{ref:holt04}
\begin{enumerate}
\item[(1)]{Okubo form}
\begin{eqnarray}
\label{eq:z-okubo}
      Z &=& P(1+\omega^{\dag}\omega )^{1/2}P,\\
V_{okb}&=& P(1+\omega^{\dag}\omega )^{1/2}P
                  (H_{0}+V_{LS})
            P(1+\omega^{\dag}\omega )^{-1/2}P
           \nonumber\\
      && -PH_{0}P.
\end{eqnarray} 
\item[(2)]{Andreozzi form}\cite{ref:andreozzi96}
\begin{eqnarray}
\label{eq:z-andre}
      Z &=& L^{T},\\
V_{andr}&=& PL^{T}P(H_{0}+V_{LS})
            P(L^{-1})^{T}P-PH_{0}P,
\end{eqnarray} 
where we put $Z^{\dag}Z = LL^{T}$ for Eq.~(\ref{eq:p-omega-z}), and  
$L$ is a lower triangle Cholesky matrix  and $L^{T}$ being its transpose.
\item[(3)]{Suzuki-Okamoto form}
\begin{eqnarray}
\label{eq:z-suzuoka}
      Z &=& P{\rm e}^{-G}{\rm e}^{\omega}P,\\
V_{suzu}&=& P{\rm e}^{-G}(H_{0}+V)
            {\rm e}^{G}P-PH_{0}P.
\end{eqnarray} 
\end{enumerate}
One can easily check the choices (\ref{eq:z-okubo}), (\ref{eq:z-andre}) and (\ref{eq:z-suzuoka}) satisfy the property (\ref{eq:p-omega-z}). 

We now can understand systematically the formal relation among
various effective hamiltonians as shown in Fig.1.
\begin{figure}[htbp]
\setlength{\unitlength}{0.8mm}
 \begin{center}
 \begin{picture}(140,135)
 \scriptsize
  \put(  0,115){\framebox(25,15)}
  \put(  0,122){ \makebox(25,7){${\cal H}(0,-1)$}}
  \put(  0,115){ \makebox(25,7){with $\chi (0,-1)$ }}
　\put( 25,120){\makebox(5,5){=}}
  \put( 32,115){\framebox(30,15)}
  \put( 32,122){ \makebox(25,7)
               {$P{\overline P}H{\overline P}P$}}
  \put( 32,115){ \makebox(27,7){with $\chi_K =P{\overline P}P$}}
    \put(10,115){\line(0,-1){7}}
    \put(44,115){\line(0,-1){7}}
    \put(10,97){\vector(0,-1){7}}
    \put(44,97){\vector(0,-1){7}}
   \put(0,95){\makebox(25,10)
        {Eliminate overlap}}
   \put( 32,95){\makebox(25,10)
        {Eliminate overlap}}
  \put( -3,75){\framebox(25,15)}
  \put( -3,75){\makebox(25,15){${\cal H}(0,0)$}}
　\put( 21,80){\makebox(5,5){=}}
  \put( 28,75){\framebox(25,15)}
  \put( 28,82){\makebox(25,7){$P{\overline P}H{\overline P}P$}}
  \put( 28,75){\makebox(25,7){$\times (P{\overline P}P)^{-1}$}}
　\put( 51,80){\makebox(5,5){=}}
  \put( 60,75){\framebox(25,15)}
  \put( 60,75){ \makebox(25,15){$PH\Omega P$}}
　\put( 85,80){\makebox(5,5){=}}
  \put( 94,75){\framebox(32,15)}
  \put( 94,75){ \makebox(30,15){$P(H_{0}+V_{LS})P$}}
    \put(  3,75){\line(0,-1){7}}
    \put( 37,75){\line(0,-1){7}}
    \put( 73,75){\line(0,-1){7}}
    \put(107,75){\line(0,-1){7}}
  \put(-10,58){\makebox(25,10){Make hermitian}}
  \put( 25,58){\makebox(25,10){Make hermitian}}
  \put( 60,58){\makebox(25,10){Make hermitian}}
  \put( 95,58){\makebox(25,10){Make hermitian}}
    \put(  3, 57){\vector(0,-1){7}}
    \put( 37, 57){\vector(0,-1){7}}
    \put( 73, 57){\vector(0,-1){7}}
    \put(107, 57){\vector(0,-1){7}}
  \put(-10,35){\framebox(25,15)}
  \put(-10,35){ \makebox(25,15)
    {${\cal H}({\scriptstyle -1/2},{\scriptstyle -1/2})$}}
　\put( 14,40){\makebox(5,5){=}}
  \put( 21,35){\framebox(25,15)}
  \put( 21,45){ \makebox(25, 5)
              {$(P{\overline P}P)^{\scriptstyle -1/2}$}}
  \put( 21,40){ \makebox(25, 5)
              {$\times (P{\overline P}H{\overline P}P)$}}
  \put( 21,35){ \makebox(25, 5)
              {$\times (P{\overline P}P)^{\scriptstyle -1/2}$}}
　\put( 46,40){\makebox(5,5){=}}
  \put( 53,35){\framebox(27,15)}
  \put( 53,42){ \makebox(25, 7)
              {$(\Omega^{\dag}\Omega)^{\scriptstyle 1/2}PH\Omega$}}
  \put( 53,35){ \makebox(25, 7)
              {$\times (\Omega^{\dag}\Omega)^{\scriptstyle -1/2}$}}
　\put( 80,40){\makebox(5,5){=}}
  \put( 87,35){\framebox(30,15)}
  \put( 87,42){\makebox(30,7){$PL^{T}P(H_{0}+V_{LS})$}}
  \put( 87,35){\makebox(30,7){$ \times P(L^{-1})^{T}P$}}
    \put(-3,35){\line(4,-3){12}}
    \put(30,35){\line(0,-1){7}}
    \put(51,27){\line(2, 1){15}}
  \put(18,18){\makebox(30,10){Convert to canonical form}}
   \put(30,20){\vector(0,-1){5}}
   \put(18,5){\framebox(25,10){$Pe^{-G}He^G P$}}
 \end{picture}
  \caption{Formal relation among various effective 
           hamiltonians in time-independent approach}
 \end{center}
\end{figure}
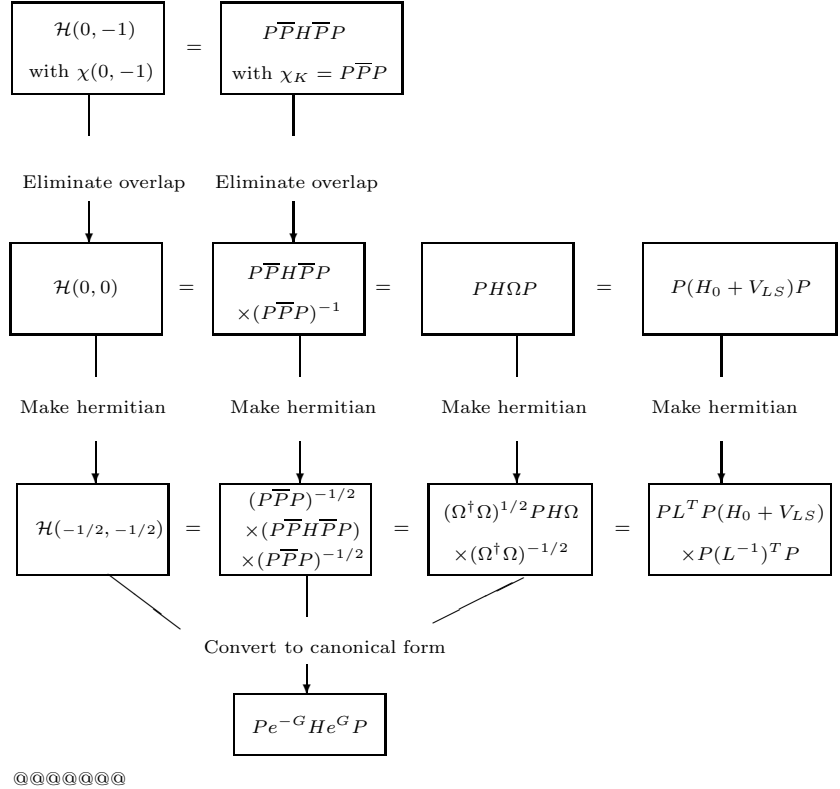
\section{Summary}
We have given the general definition of the $P$-space 
effective hamiltonian or interaction, mainly  the $E$-independent
effective hamiltonians derived in a time-independent way.
 We have discussed the relation among hermitian and non-hermitian 
effective hamiltonians.
\section*{Acknowledgements}
One of the authors (R.O.) thanks James Vary for his continuous interest
on the formal relation between various effective interactions.

He is very grateful to the organizers for giving him to speak
on this subject.
 He also thanks the  Feza Gursey institute for warm hospitality.
This work was supported by a Grant-in-Aid for Scientific Research (C) from
Japan Society for the Promotion of Science (JSPS) (Grant No. 15540280).


\end{document}